\documentclass[12pt]{iopart}
\usepackage{amssymb}
\usepackage{graphicx}
\usepackage{amssymb}
\usepackage{dsfont}
\usepackage{tensor}
\usepackage{upgreek}
\usepackage{xcolor}
\usepackage{braket}
\usepackage{upgreek}

\begin{document}
\pdfminorversion=4
\title[Effect of phonons on the electron spin resonance absorption spectrum]
{Effect of phonons on the electron spin resonance absorption spectrum}

\author{Ariel Norambuena$^{1}$, Alejandro Jimenez$^{2}$, Christoph Becher$^{3}$ and Jer\'onimo R. Maze$^{2}$}

\address{$^1$ Centro de Investigaci\'on DAiTA LAB, Facultad de Estudios Interdisciplinarios, Universidad Mayor, Santiago, Chile}
\address{$^2$ Faculty of Physics, Pontificia Universidad Cat\'olica de Chile, Avda. Vicu\~{n}a Mackenna 4860, Santiago, Chile}
\address{$^3$ Universit\"{a}t des Saarlandes, Fachrichtung Physik, Campus E2.6, 66123 Saarbr\"{u}cken, Germany}

\ead{ariel.norambuena@umayor.cl, jeromaze@gmail.com}
\vspace{10pt}

\begin{abstract}

The unavoidable presence of vibrations in solid-state devices can drastically modify the expected electron spin resonance (ESR) absorption spectrum in magnetically active systems. In this work, we model the effect of phonons and temperature on the ESR signal in molecular systems with strong $E \otimes e$ Jahn-Teller (JT) effect and an electronic spin-$1/2$. Our microscopic model considers the linear JT interaction with a continuum of phonon modes, the spin-orbit coupling, the Zeeman effect, and the response of the system under a weak oscillating magnetic field. We derive a Lindblad master equation for the orbital and spin degrees of freedom, where one- and two-phonon processes are considered for the phonon-induced relaxation, and the thermal dependence of Ham reduction factors is calculated. We find that the suppression of ESR signals is due to phonon broadening but not based on the common assumption of orbital quenching. Our results can be applied to explain the experimentally observed absence of the ESR signal in color centers in diamond, such as the neutral nitrogen-vacancy and negatively charged silicon-vacancy color centers in diamond.

\end{abstract}

%
%
\maketitle
%
%

\section{Introduction}
Since its discovery in 1945~\cite{Zavoisky1945}, electron paramagnetic resonance (EPR) or electron spin resonance (ESR) has been extensively used to characterize molecular systems~\cite{Weil1986}, quantum dots~\cite{Xiaojie2014}, metal complexes~\cite{Dyson1955}, organic radicals~\cite{Gerson2004} or defects in solid-state systems~\cite{Richelle2016}. Commonly, ESR consists of applying a constant frequency microwave field while a magnetic field is swept across spin-flip resonance transitions. Crucial information has been obtained in several biochemical problems~\cite{Sahu2013}, on Hyperfine interactions with high resolution~\cite{Piette1967} and on magnetic ions in metals~\cite{Taylor1975}, to name a few examples. On the other hand, by analyzing the ESR absorption spectrum and dispersion lines, effects such as anisotropy of gyromagnetic factors~\cite{Malissa2004}, Ham reduction factors~\cite{Ham1965, Ham1968}, and spin relaxation rates~\cite{Damon1953} have been studied. \par

In many systems were a paramagnetic behavior is expected, however, no ESR absorption is detected~\cite{Felton2008}. Especially on those systems that are strongly coupled to vibrations~\cite{Davies1979}, the ESR response is not observed at high temperatures. Particularly, experimental measurements with defects in diamond reveal an absence of the ESR signal at room temperature, which is observed for the negatively charged silicon-vacancy (SiV$^{-}$)~\cite{Edmonds2008} and the neutral nitrogen-vacancy (NV$^{0}$) centers in diamond~\cite{Felton2008}. In both cases, spin $S=1/2$ quantum systems dominated by the dynamical Jahn-Teller (JT) effect are quoted as examples of ESR suppression due to the fast orbital averaging effect (Ham reduction factors), but with no mention of a theoretical explanation. Here, we demonstrate that a strong dynamical JT effect primarily modifies the orbital states through the Ham reduction factors, which changes the energy of the resonant transitions of the system instead of the amplitude of the ESR signal. Only by including phonon-induced relaxation processes we reproduce the observed suppression of the ESR signal at room temperature, which is conclusively explained by the phonon broadening effects induced by two-phonon Raman processes. \par

Although many works have been devoted to phenomenologically describing the ESR response, still few of them deal with microscopic aspects of the electron-phonon coupling and its thermal effects. Several authors have addressed the problem of modeling the ESR response using out-of-equilibrium models~\cite{Kittel1948, Anderson&Weiss1953, Kubo1954}. Historically, the linear response theory associated with irreversible processes was introduced in order to give a more satisfactory description of the magnetic resonance phenomenon~\cite{Kubo1954}. Nevertheless, the inclusion of thermal effects induced by electron-phonon processes on the ESR response has not yet been microscopically solved in trigonal systems with a strong $E \otimes e$ JT effect. In this work, we present a microscopic model for the ESR response in a generic $E \otimes e \otimes SU(2)$ system, where we include the spin-$1/2$ degree of freedom. Such a system can be, for instance, the NV$^{0}$ and the SiV$^{-}$ centers, or more general: the negatively charged group IV-vacancy centers. Furthermore, we derive a thermal dependence of Ham reduction factors, which is consistent with recent \textit{ab initio} calculations for defects in diamond~\cite{Gergo2018} (zero temperature) and the original theory presented by Ham~\cite{Ham1965, Ham1968}. \par

This paper is organized as follows. In section~\ref{Hamiltonian-of-the-system}, we introduce the Hamiltonian of the $E \otimes e \otimes SU(2)$ system, where a large band gap between the ground and first-optically excited state is assumed. In section~\ref{Effective dynamics and Ham reduction factors}, we derive an effective Hamiltonian for the orbital and spin degrees of freedom induced by the strong linear JT coupling. Furthermore, we introduce an analytical expression for the thermal dependence of Ham reduction factors, where the zero-temperature case is obtained from recent \textit{ab initio} calculations. In section~\ref{Lattice phonons and quantum master equation}, we introduce the phonon-induced relaxation processes using a Lindblad master equation that includes one- and two-phonon processes. In section~\ref{ESR absorption spectrum}, we use the linear response theory, and we find an expression for the ESR absorption spectrum in terms of the resonant frequencies and the phonon relaxation rates, where a super-Ohmic environment is assumed. Finally, we discuss the suppresion of the ESR signal, and in section~\ref{Conclusion}, we summarize our results. \par

\section{Hamiltonian of the system}\label{Hamiltonian-of-the-system}

Let us consider a trigonal system with $D_{3}$ symmetry whose degenerate orbital states $\ket{X}$ and $\ket{Y}$ with symmetry $E$ are coupled to multiple $e$-phonon modes. Traditionally this type of orbital-phonon interaction is modeled using the $E \otimes e$ JT theory~\cite{Bersuker2006}. Now, we shall include an internal spin degree of freedom $S = 1/2$ to model the effect of phonons on the ESR signal. Towards this end, we incorporate the spin states $\ket{\uparrow} = \ket{m_s = +1/2}$ and $\ket{\downarrow} = \ket{m_s=-1/2}$ into the dynamics of the system. In general, under the presence of a constant $\mathbf{B}$ and weak time-dependent $\mathbf{b}(t)$ magnetic fields, the Hamiltonian of the $E \otimes e \otimes SU(2)$ system is given by 

\begin{eqnarray} \label{TotalHamiltonian}
\hat{H} =  \hat{H}_{\rm JT} + \hat{H}_{\rm so}+  \hat{H}_{\rm z}+ 
\hat{H}_{\rm strain} + \hat{V}(t).
\end{eqnarray}

For multiple phonon modes the $E \otimes e$ JT Hamiltonian is defined as~\cite{Bersuker2006}

\begin{eqnarray} \label{JTHamiltonian} 
\hat{H}_{\rm JT} = E_0 + \sum_{k \in E} F_k (\hat{Q}_{k \rm x} \hat{\sigma}_z - \hat{Q}_{k \rm y} \hat{\sigma}_x) +\sum_{k \in E,i= \rm x,\rm y}\left({\hat{P}_{k i}^2 \over 2\mu_k} + {1 \over 2}\omega_k^2\hat{Q}_{k i}^2\right),
\end{eqnarray} 

where $E_0$ is the energy of the degenerate states $|X,\uparrow\rangle$, $|X,\downarrow\rangle$, $|Y,\uparrow\rangle$ and $|Y,\downarrow\rangle$. The last two terms of the right-hand of Hamiltonian~(\ref{JTHamiltonian}) describe the linear electron-phonon coupling and the energy of the harmonic oscillators. Here, $k \in E$ means that we are considering phonon modes belonging to the irreducible representations $E$, and in what follow we called $e$-phonon modes. The term $F_k$ is the linear vibronic coupling constant, while $\mu_k$ and $ \omega_k$ are the reduced mass and the phonon frequencies, respectively. In the orbital manifold $\{\ket{X},\ket{Y}\}$, the operators are defined as $\hat{\sigma}_x = |X\rangle \langle Y| + |Y\rangle \langle X|$, $\hat{\sigma}_y = -i|X\rangle \langle Y| + i|Y\rangle \langle X|$ and $\hat{\sigma}_z = |X\rangle \langle X| - |Y\rangle \langle Y|$. The second term in equation~(\ref{TotalHamiltonian}) is the spin-orbit interaction 

\begin{eqnarray} \label{SOHamiltonian}
\hat{H}_{\mbox{\scriptsize so}} = -\lambda (\mathbf{L} \cdot \mathbf{S}) = -\lambda \hat{L}_{\rm z}  \hat{S}_{\rm z}, 
\end{eqnarray}

\begin{figure}[htb]
\centerline{\includegraphics[width=0.65\textwidth]{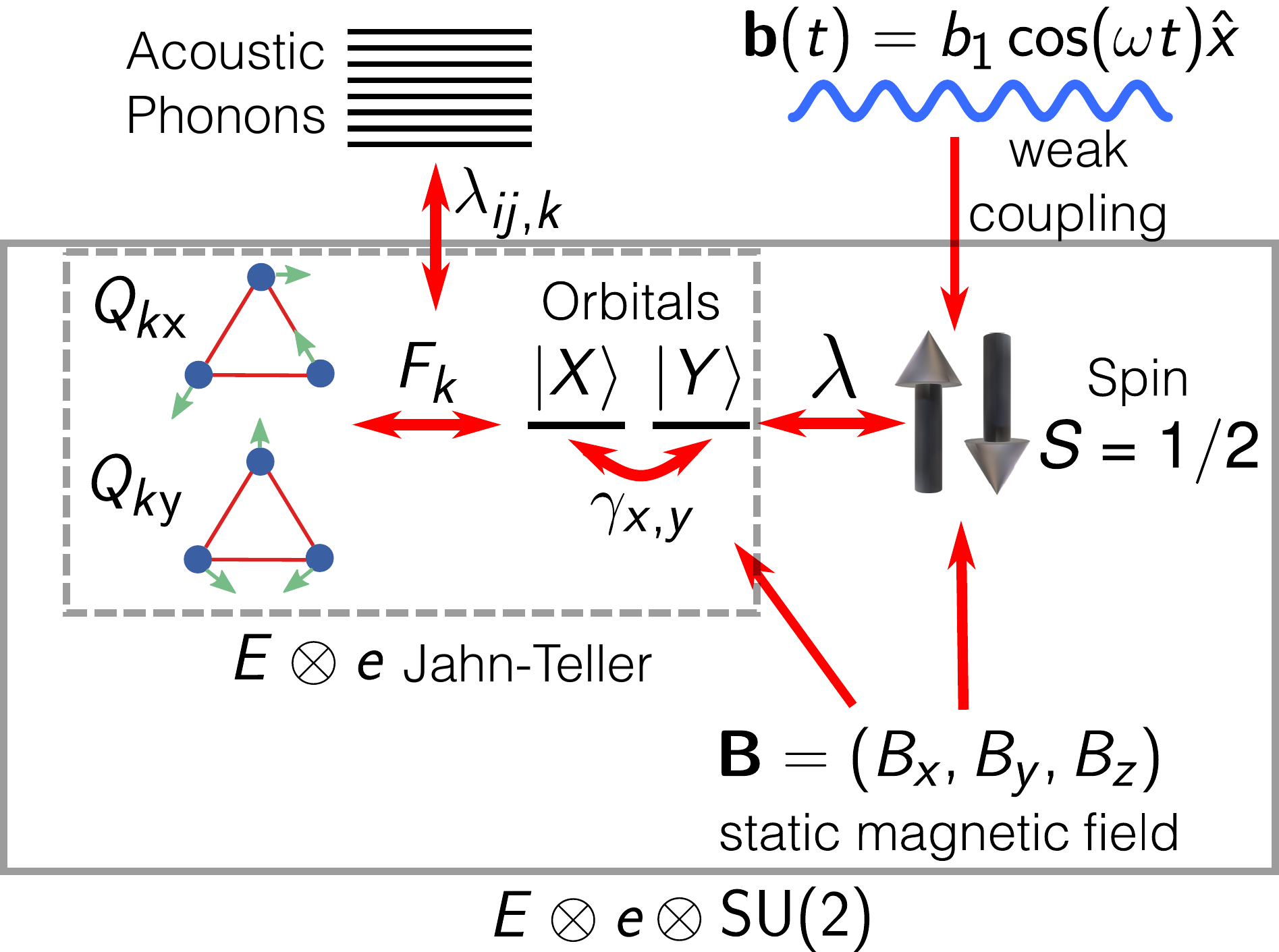}}
\caption{Schematic representation of the $E \otimes e \otimes \mbox{SU(2)}$ system perturbed by an oscillating magnetic field. The dashed box enclosed the usual $E \otimes e$ Jahn-Teller system, which is coupled to the spin states via the spin-orbit coupling constant $\lambda$. The static magnetic field interacts with both orbital and spin degrees of freedom, while the oscillating magnetic field only perturbs the spin states. Resonant acoustic phonons are depicted as a thermal bath interacting with the orbital states.} \label{figure1}
\end{figure}

where $\lambda > 0$ is the spin-orbit coupling constant, $\mathbf{L}$ is the orbital angular momentum operator, and $\mathbf{S} = {1 \over 2}(\hat{s}_{\rm x},\hat{s}_{\rm y},\hat{s}_{\rm z})$. The spin operators are given by $\hat{s}_x = \ket{\uparrow}\bra{\downarrow}+\ket{\downarrow} \bra{\uparrow}$, $\hat{s}_y = -i\ket{\uparrow}\bra{\downarrow}+i\ket{\downarrow} \bra{\uparrow}$ and $\hat{s}_z = \ket{\uparrow}\bra{\uparrow}-\ket{\downarrow} \bra{\downarrow}$. In the term $-\lambda \hat{L}_{\rm z}  \hat{S}_{\rm z}$ we are assuming that orbital states with symmetry different from $E$ are far in energy from the doubly degenerate $E$-states. As a consequence, we neglect the effect of operators $\hat{L}_{\rm x}$ and $\hat{L}_{\rm y}$ in our manifold since these orbital operators mix orbital states separated by a large band gap (ground and excited states). Additionally, we have $\hat{L}_{\rm z} = \hbar \hat{\sigma}_{\rm y}$ which is a useful representation for the $z$-component of the angular momentum in the orbital basis $\{\ket{X},\ket{Y}\}$. The third term in equation~(\ref{TotalHamiltonian}) is the spin and orbital Zeeman Hamiltonian associated with the static magnetic field $\mathbf{B} = (B_{\rm x},B_{\rm y},B_{\rm z})$ ($\hbar = 1$)

\begin{eqnarray}\label{ZeemanHamiltonian}
\hat{H}_{\mbox{\scriptsize z}} &=  \left(\gamma_{\rm s} \mathbf{S} + \gamma_{\rm L} \mathbf{L} \right) \cdot \mathbf{B} \approx {1 \over 2}\gamma_{\rm s} \sum_{i= \rm{x,y,z}} B_i \hat{s}_i +{1 \over 2} \gamma_{\rm L} B_{\rm z} \hat{\sigma}_{\rm y}, 
\end{eqnarray}

where $\gamma_{\rm s} = \mu_{\rm B} g_{\rm s}/\hbar$ and $\gamma_{\rm L} = \mu_{\rm B} g_{\rm L}/ \hbar$ are the spin and orbital gyromagnetic constants, respectively, with $\mu_{\rm B}$ being the Bohr magneton. Here, $g_{\rm s} = 2$ and $g_{\rm L} = 1$ are the Land\'e $g$-factors. Since we have neglected the operators $\hat{L}_{\rm x}$ and $\hat{L}_{\rm y}$ it follows that $ \mathbf{L} \cdot \mathbf{B} \approx \hat{L}_{\rm z} B_{\rm z}$, which justifies the approximation on equation~(\ref{ZeemanHamiltonian}). The fourth term in equation~(\ref{TotalHamiltonian}) is the strain Hamiltonian for a trigonal system~\cite{Bersuker2006, Hepp2014}

\begin{eqnarray} \label{StaticStrain}
\hat{H}_{\rm strain} = {1 \over 2}\left(\gamma_{\rm x} \hat{\sigma}_{\rm z} - \gamma_{\rm y} \hat{\sigma}_{\rm x}\right), 
\end{eqnarray}

where $\gamma_{\rm x}$ and $\gamma_{\rm y}$ are the coupling constants that describe the effect of static distortions. Finally, the last term on equation~(\ref{TotalHamiltonian}) is the time-dependent Hamiltonian associated with the weak oscillating magnetic field $\mathbf{b}(t) = b_1 \cos (\omega t)\mathbf{e}_x$ 

\begin{equation}\label{Perturbation}
\hat{V}(t) = \gamma_{\rm s} \mathbf{S} \cdot \mathbf{b}(t)  = \gamma_{\rm s} b_1  \hat{S}_{\rm x} \cos(\omega t),
\end{equation}

Figure~\ref{figure1} shows a schematic illustration of the $E\otimes e \otimes \mbox{SU(2)}$ system, illustrating the dynamics presented in our model. In the next sections, we derive the effective dynamics for the orbital and spin degrees of freedom, and we include the effect of a phononic bath composed of acoustic phonons.

\section{Effective dynamics for the orbital and spin degrees of freedom}\label{Effective dynamics and Ham reduction factors}

In this section, the important case of color centers in diamond is analysed to emphasizes the role of a strong JT effect on the orbital and spin dynamics. First, we focus on the characteristic time scales of the JT effect in comparison with the phonon-induced relaxation processes, which will be our most important starting point. The JT effect has been numerically studied using first-principles calculations for the $^3E$ excited state of the NV$^{-}$ center~\cite{Abtew2011}, both excited and ground states of the SiV$^{0}$~\cite{NatureAdam2019} and SiV$^{-}$ centers~\cite{Gergo2018}. We focus our analysis to the particular case of the SiV$^{-}$ center in diamond to illustrate the main ideas of the microscopic model. For the SIV, the numerically estimated JT and barrier energies are $E_{\rm JT} = 42.3$ meV and $\delta_{\rm JT} = 3.0$ meV, respectively~\cite{Gergo2018}. Using the previous values and the relations $E_{\rm JT} = F^2/(2/(\hbar \omega-2G))$ and $\delta_{\rm JT} = 4 E_{\rm JT} G/(\hbar \omega + 2G)$ it follows that the linear vibronic JT coupling is $F = 83.34$ meV, which is associated to a local vibrational mode with energy $\hbar\omega = 85.2$ meV. The latter implies that the time scale associated with the JT interaction is $\tau_{\rm JT} \sim g_{\rm JT}^{-1} \approx 0.3$ ps, which is a fast dynamics. We expect a similar time scale for the JT dynamics (few picoseconds) for the ground state of the neutral nitrogen-vacancy center. On the other hand, experimental measurements of the phonon-induced relaxation rates show that $\Gamma_{\rm ph} \sim (10^{-5}-10^7)$ s$^{-1}$ for temperature ranging from few mK to room temperature~\cite{Jarmola2012, Goldman2015, Aster2018, Ariel2018, Gugler2018,Becker2016}, where a microscopic model for the electron-phonon processes of the SiV$^{-}$ center is discussed in Ref.~\cite{Janke2015}. As a consequence, phonon-induced relaxation processes are relevant for times larger than $\tau_{\rm ph-relax} \sim 0.1$ $\upmu$s. Therefore, in any physical system satisfying the condition $\tau_{\rm JT} \ll \tau_{\rm ph-relax}$ the strong JT effect modifies the energy levels of the system in the short-time dynamics, and the phonon-induced relaxation processes will be resonant to an effective Hamiltonian dressed by the JT effect, which is the original idea of the Ham reduction factors~\cite{Ham1965,Ham1968}. \par

In the time scale $0 < t \sim \tau_{\rm JT}$, and assuming that $\tau_{\rm JT} \ll \tau_{\rm ph-relax}$, the system is ruled by the following Liouville–von Neumann equation

\begin{eqnarray} \label{EffectiveLiouvilleEquation}
{\mbox{d} \hat{\rho}_{\rm t} \over \mbox{d} t} = {1 \over i\hbar}[\hat{H}_{\rm t},\hat{\rho}_{\rm t}], \quad \quad \hat{H}_{\rm t} \approx \hat{H}_{\rm JT} + \hat{H}_{\rm so} +  \hat{H}_{\rm z}+ \hat{H}_{\rm strain},
\end{eqnarray}

where $\hat{\rho}_{\rm t}$ is the density matrix associated with the orbital, spin, and $e$-phonons degrees of freedom. Here, the time-dependent magnetic field is not considered as it has a much slower dynamics. In equation~(\ref{EffectiveLiouvilleEquation}), we have an approximate Hamiltonian because we are neglecting the electron-phonon interaction responsible for the phonon-induced relaxation processes. Now, we will focus our analysis on the case of strong JT interaction. In such a case, the Hamiltonian $\hat{H}_{\rm JT}$ is not a small perturbation, and therefore the usual weak-coupling approximation of the master equation cannot be applied to derive the dynamics for the orbital and spin states. An alternative approach is to find a new reference frame or unitary transformation in which the coupling with $e$-phonons will be small, and then we could apply perturbation theory. To accomplish the previous observation, we introduce the following unitary transformation

\begin{eqnarray} \label{Transformation}
\hat{H}_{\rm t}' = \hat{U} \hat{H}_{\rm t} \hat{U}^{\dagger}, \quad \quad  \hat{U} = \mbox{exp}\left(i \sum_{k \in E} \epsilon_k \hat{T}_k \right),
\end{eqnarray}

where $\hat{T}_k = \hat{P}_{k \rm x} \hat{\sigma}_z+\hat{P}_{k \rm y} \hat{\sigma}_{\rm x}$ is a linear phonon-orbital operator and $\epsilon_k = -F_k/\mu_k\hbar \omega_k$, where $F_k$, $\mu_k$ and $\omega_k$ are the parameters introduced in the JT Hamiltonian~(\ref{JTHamiltonian}). The summation introduced in equation~(\ref{Transformation}) only considers the contribution of $e$-phonon modes responsible of the JT effect. The transformation~(\ref{Transformation}) was originally introduced by Ham in his pioneering work of the quenching effect of the orbital operators in $E \otimes e$ systems with strong JT interaction~\cite{Ham1968}. The coefficient $\epsilon_k \sim F_k/\hbar \omega_k$ can be understood as the ratio between the vibronic coupling and the phonon energy, and for color centers in diamond this factor is small ($\epsilon \approx 0.1$ for the SiV$^{-}$ center). In the regime $F_k/\mu_k \hbar \omega_k \ll 1$, we obtain

\begin{eqnarray}
\hat{H}_{\rm t}' \approx 
E_0 + \sum_{k \in E,i= \rm x,\rm y}\left({\hat{P}_{k i}^2 \over 2\mu} + {1 \over 2}\omega_k^2\hat{Q}_{k i}^2\right) - \sum_{k \in E}{F_k^2 \over \mu_k \hbar \omega_k^2}\left(3\hbar - 4\hat{L}_z \hat{L}_{k \rm z}^{\rm ph}\right).
\end{eqnarray}

The first two terms of the right-hand side describe the free energy of the degenerate orbital states and $e$-phonon modes, respectively. More importantly, the last term defines an exchange of angular momentum between orbital states and $e$-phonons. The orbital and phonon momentum operators are defined as $\hat{L}_z =  \hbar \hat{\sigma}_{\rm y}$ and $\hat{L}_{k \rm z}^{\rm ph} = \hbar^{-1}(\hat{P}_{k \rm x} \hat{Q}_{k \rm y}-\hat{P}_{k \rm y} \hat{Q}_{k \rm x})$, respectively. After applying the transformation~(\ref{Transformation}) the Liouville–von Neumann equation~(\ref{EffectiveLiouvilleEquation}) transform as $\dot{\hat{\rho}}_{\rm t}' = (i\hbar)^{-1}[\hat{H}_{\rm t}',\hat{\rho}_{\rm t}']$ with $\hat{\rho}_{\rm t}' = \hat{U} \hat{\rho}_{\rm t} \hat{U}^{\dagger}$. To leading order in $\epsilon_k$ and calculating $\hat{\rho}_{\rm s} = \mbox{Tr}_{\rm ph}(\hat{\rho}_{\rm t}')$, we obtain the following effective closed dynamics for the orbital and spin degrees of freedom

\begin{equation}\label{EffectiveDynamics}
{\mbox{d} \hat{\rho}_{\rm s} \over \mbox{d} t} = {1 \over i\hbar} [\hat{H}_{\rm eff}, \hat{\rho}_{\rm s}].  
\end{equation}  

The most important result is the effective Hamiltonian 

\begin{eqnarray} \label{EffectiveHamiltonian}
\hat{H}_{\rm eff} &= E_0 - p \lambda \hat{S}_{\rm z} \hat{L}_{\rm z} + \gamma_{\rm s} \mathbf{S} \cdot \mathbf{B} + p \gamma_{\rm L} \hat{L}_{\rm z} B_{\rm z} +  {q \over 2}\left(\gamma_{\rm x} \hat{\sigma}_{\rm z} - \gamma_{\rm y} \hat{\sigma}_{\rm x}\right),
\end{eqnarray}

\begin{figure}[htb]
\centerline{\includegraphics[width=0.8\textwidth]{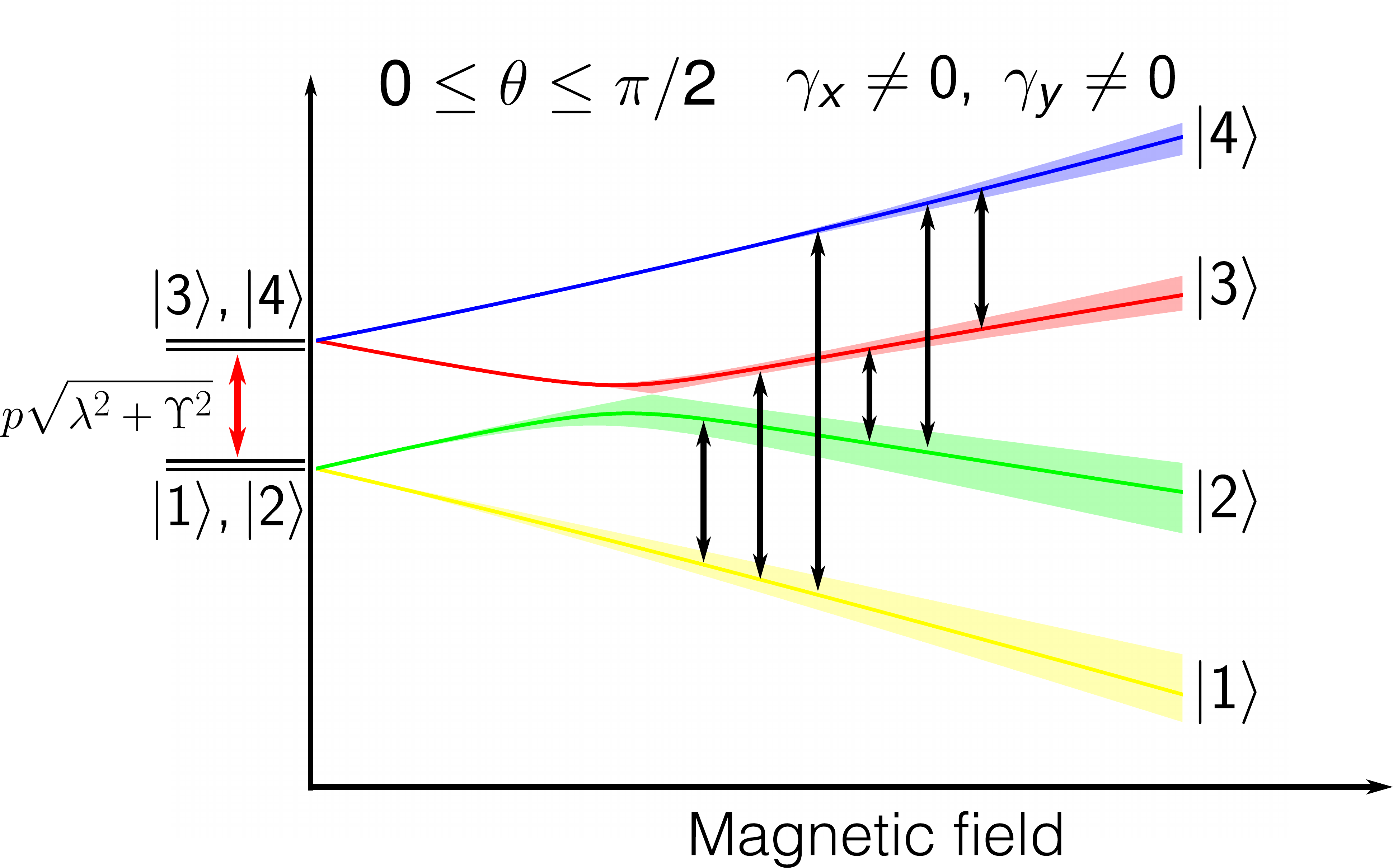}}
\caption{Energy levels of the $E \otimes e \otimes SU(2)$ system as a function of the magnitude of the magnetic field $\mathbf{B}$ for a fixed temperature. The filled coloured areas represent the possible values of the energies $E_i$ for $0 \leq \theta \leq \pi/2$, $\lambda = 50$ GHz and $\gamma_{x,y} = 2$ GHz. The transitions between the states are illustrated with black arrows which can be realized by one- and two-phonon processes.
} \label{figure2}
\end{figure}

where $p$ and $q$ are the Ham reduction factors which satisfy the relation $q = (1+p)/2$~\cite{Ham1968}. In figure~\ref{figure2} we illustrate the energy levels of the SiV$^{-}$ center as a function of the magnetic field $\mathbf{B} = B(\cos \phi \sin \theta \mathbf{e}_x+\sin \phi \sin \theta \mathbf{e}_y + \cos \theta \mathbf{e}_z)$, where the energy gap $p\sqrt{\lambda^2+\Upsilon^2}$ is affected by the Ham reduction factor $p$, with $\Upsilon = \sqrt{\gamma_{\rm x}^2+\gamma_{\rm y}^2}$ being the total strain coupling. In the weak coupling regime ($\epsilon_k \ll 1$) and to first-order, the Ham reduction factor is reduced to 

\begin{equation}\label{p-factor}
p^{(1)} = 1 - 2 \sum_{k \in E} {|g_{k}|^2 \over (\hbar \omega_k)^2 }\coth\left(\hbar \omega_k \over 2 k_{\rm B} T \right),
\end{equation}

where $g_{k} = F_k \sqrt{\hbar/2\mu \omega_k}$, $k_{\rm B}$ is the Boltzmann constant and $T$ is the reservoir temperature. For a detailed derivation of the Ham reduction factor see~\ref{AppendixA}. \par

\subsection{Thermal dependence of the Ham reduction factor}
To calculate the thermal dependence of the Ham reduction factor $p$ for a continuum of $e$-phonon modes is necessary to consider corrections beyond the first-order approximation given in equation~(\ref{p-factor}). In this direction, second-order corrections lead to~\cite{Napoleon} 

\begin{eqnarray}\label{p-factor2}
p^{(2)} = 1 - 2 \sum_{k \in E} {|g_{k}|^2 \over (\hbar \omega_k)^2 }\coth\left(\hbar \omega_k \over 2 k_{\rm B} T \right) + 4\left[ \sum_{k \in E} {|g_{k}|^2 \over (\hbar \omega_k)^2 }\coth\left(\hbar \omega_k \over 2 k_{\rm B} T \right)  \right]^2.
\end{eqnarray}

From equations~(\ref{p-factor}) and (\ref{p-factor2}) we observe that the Taylor expansion of the Ham reduction factor has the form $p = 1-2x+4x^2 + ... = \mbox{exp}(-2x)$ with $x = \sum_{k \in E} |g_{k}|^2 / (\hbar \omega_k)^2 \coth (\hbar \omega_k / 2 k_{\rm B} T)$. Therefore, when $e$-phonons are considered as a continuum field, we obtain the following analytical expression 

\begin{equation} \label{HamReductionFactor}
p =  \mbox{exp}\left[- 2 \int_{0}^{\infty} {J(\omega) \over (\hbar \omega)^2} \coth\left({\hbar \omega \over 2 k_{\rm B} T} \right) \, d\omega \right],
\end{equation}

where $J(\omega)$ is the phononic spectral density function associated with $e$-phonon modes

\begin{equation}
J(\omega) = \sum_{k \in E} |g_k|^2 \delta(\omega-\omega_k).
\end{equation}

In principle, the spectral density function takes into account the contribution of all $e$-phonon modes associated with the JT effect and must be numerically calculated. Recent \textit{ab initio} calculations show no indication of the presence of quasi-localized phonon modes with symmetry $e_g$ in the ground state of the SiV$^{-}$ center~\cite{Londero2018}. Assuming the absence of distinct quasi-localized resonances in the shape of the phononic spectral density function $J(\omega)$, we expect a dominant contribution coming from lattice phonons. Therefore, as a first theoretical approach, we consider the following super-Ohmic model for the phononic spectral density function 

\begin{equation} \label{SpectralDensityFunction}
J_{\rm acous}(\omega) = \alpha  \omega^3 e^{-\omega/\omega_{\rm c}},
\end{equation}

where $\alpha = \Omega g_{\rm JT}^2/(2 \pi^2 v_{\rm s}^3 \omega_{\rm D})$ and $\omega_{\rm c} \simeq 2$ THz is the cut-off frequency for the acoustic phonons in diamond. For a diamond lattice we have $\Omega = a^3$, where $a = 3.57 \; \mbox{\AA}$ is the lattice constant, $v_{\rm s} = 1.2 \times 10^4$ m/s is the speed of sound, $\omega_{\rm D} \approx 40 \times 10^{12}$ Hz is the Debye frequency, and $g_{\rm JT}$ is the electron-phonon coupling constant. By direct integration of equation~(\ref{HamReductionFactor}) using the spectral density function $J_{\rm acous}(\omega)$~(\ref{SpectralDensityFunction}), we obtain 

\begin{equation}
p_{\rm acous}(T) = \mbox{exp}\left[-{\Omega g_{\rm JT}^2 \omega_{\rm c}^2 \over \pi^2 v_{\rm s}^3 \hbar^2 \omega_{\rm D}} \left(2\left({k_{\rm B} T \over \hbar \omega_{\rm c}}\right)^2\psi^{(1)}\left({k_{\rm B} T \over \hbar \omega_{\rm c}}\right)-1\right) \right],
\end{equation}

where $\psi^{(m)}(x) = (-1)^{m+1}m!\sum_{k=0}^{\infty}(x+k)^{-(m+1)}$ is the polygamma function. At zero temperature, we obtain $p_{\rm acous}(0) = \mbox{exp}[- \Omega g_{\rm JT}^2 \omega_{\rm c}^2 /(\pi^2 v_{\rm s}^3 \hbar^2 \omega_{\rm D})]$. Numerical calculations of the Ham reduction factor at zero temperature for the SiV$^{-}$ center in diamond leads to $p_{\rm num} \approx 0.308$~\cite{Gergo2018}. By comparison we deduce that $g_{\rm JT}/2\pi = (v_{\rm s}^3 \hbar^2 \omega_{\rm D} \ln(p_{\rm num}^{-1})/(4\Omega \omega_{\rm c}^2))^{1/2}$ which can be used to calculate the electron-phonon coupling associated with $e$-phonons for a fixed set of parameters $(\Omega,\omega_{\rm c},v_{\rm s},p_{\rm num})$.  \par

\begin{figure}[htb]
\centerline{\includegraphics[width=0.6\textwidth]{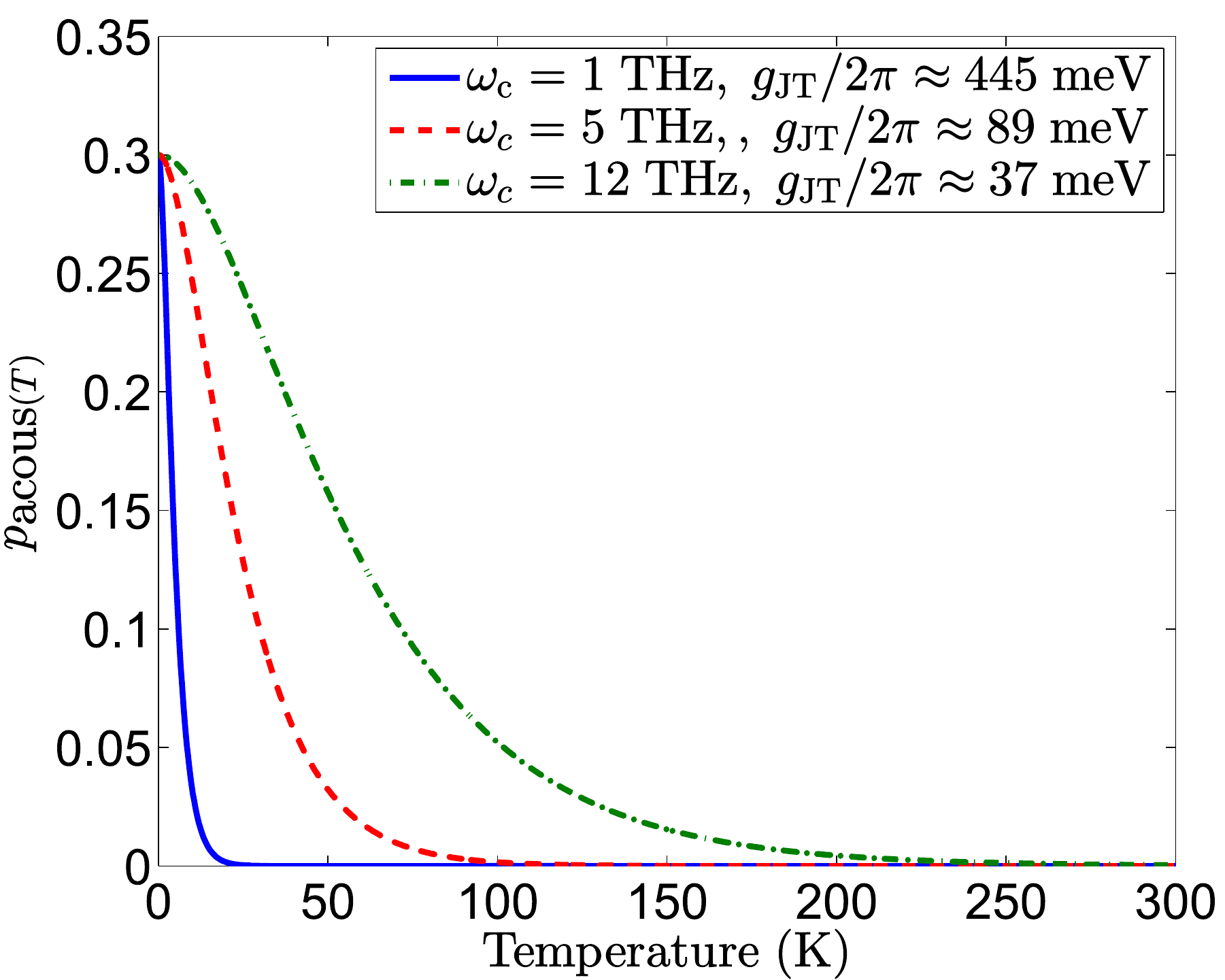}}
\caption{Thermal dependence of the Ham reduction factor $p_{\rm acous}(T)$ for different cut-off frequencies $\omega_{\rm c}$. The three curves start with a initial value of $p_{\rm acous}(0) = 0.3$ but for different cut-off frequencies we observe different decays. For the higher cut-off frequency $\omega_{\rm c} = 12$ THz (dot-dashed line) the contribution of the spectral density function $J(\omega) = \alpha \omega^3 \exp(-\omega/\omega_{\rm c})$ into the integral ~(\ref{HamReductionFactor}) takes smaller values in the region of low energy phonons ($0<\omega<5$ THz) leading to a slow varying dependence of the Ham reduction factor. In contrast, for smaller values of $\omega_{\rm c}$ (dashed and solid lines) the decay of the factor $p$ is faster.} \label{figure3}
\end{figure}

The cut-off frequency $\omega_{\rm c}$ is a lattice-dependent parameter, and therefore it is instructive to model the expected thermal dependence of $p_{\rm acous}(T)$ for systems having a different $\omega_{\rm c}$ but equal initial value $p(0) \approx 0.3$. In figure~\ref{figure3} we show the thermal dependence of the Ham reduction factor for different situations that fulfil the condition $p_{\rm acous}(0) = 0.3$. We observe that the contribution of multiple $e$-phonon modes can drastically decrease the Ham reduction factor. This is one of the most important results of this work, and we will study further implications of this behavior on the ESR signal. It is important to observe what happens in the high-temperature regime and weak static strain, \textit{i.e.} $\gamma_{\rm x,y} \ll \lambda$.

This is the case for color centers in diamond with a small strain induced by the crystal distortion. At high temperatures ($T \sim 300$ K) the Ham reduction factor $p$ in equation~(\ref{EffectiveHamiltonian}) eliminates the contribution of the spin-orbit interaction leading to an effective Hamiltonian $\hat{H}_{\rm eff} \approx E_0  + \gamma_{\rm s} \mathbf{S} \cdot \mathbf{B}$. Thus, under these approximations, in an ESR experiment at room temperature, the resonances will be centered around the Zeeman frequencies. However, under realistic experimental conditions, the peaks of the ESR signal are broadened by the phonon relaxation rates. In the next section, we introduce the effect of phonon relaxation in order to quantify the role of one- and two-phonon processes on the ESR absorption spectrum. \par

\section{Open dynamics and phonon relaxation processes}\label{Lattice phonons and quantum master equation}

To include the phonon-induced relaxation processes, we consider the dynamics in the time scale $t \gg \tau_{\rm JT }$. For color centers in diamond, the contribution of one- and two-phonon processes are crucial to model the thermal dependence of the phonon relaxation rates and their effects on the ESR signal. From experiments and theory is known that the most relevant phonon relaxation processes observed in color centers are the following: i) direct processes with the energy condition $\hbar \omega_k = E_i-E_j$, and ii) two-phonon processes with $\hbar(\omega_k -\omega_{k'})= E_i-E_j$~\cite{Jarmola2012,Janke2015,Green2017}. In our model, the energy condition of the previous processes fully matches the energy gap $E_i-E_j$ between different eigenstates of the effective Hamiltonian~(\ref{EffectiveHamiltonian}), where $\omega_k$ are the acoustic phonon frequencies of the diamond lattice. As a consequence of the strong JT effect, the relaxation processes start to be observable in a time scale where the orbital reduction (Ham reduction factors) has been reached. Formally, the relaxation rates can be derived from the electron-phonon interaction and employing the Fermi golden rule theory~\cite{Ariel2018}. In the basis spanned by the eigenstates of the effective Hamiltonian~(\ref{EffectiveHamiltonian}), we can write the electron-phonon Hamiltonian for one- and two-phonon processes as follows: 

\begin{eqnarray} \label{eph}
\hat{H}_{\rm e-ph} &=& \sum_{i\neq j}^{} \sum_{k:\; \omega_k = v_{\rm s} |\mathbf{k}|}\lambda_{ij,k}|i\rangle \langle j|(\hat{b}_k^{\dagger}+\hat{b}_k) \nonumber \\
&&+\sum_{i\neq j}^{} \sum_{k: \; \omega_k = v_{\rm s} |\mathbf{k}|} \; \sum_{k': \; \omega_{k'} = v_{\rm s} |\mathbf{k}'|}\lambda_{ij,kk'}|i\rangle \langle j|(\hat{b}_k^{\dagger}+\hat{b}_k)(\hat{b}_{k'}^{\dagger}+\hat{b}_{k'}),  
\end{eqnarray}

where phonons with a linear dispersion relation $\omega_k = v_s |\mathbf{k}|$ are considered in the electron-phonon Hamiltonian, being $v_s$ the speed of sound and $\mathbf{k}$ the wavevector for phonons. Here, $\lambda_{ij,k}$ and $\lambda_{ij,kk'}$ are the linear and quadratic electron-phonon coupling constants, respectively and $\hat{b}_k^{\dagger}$ ($\hat{b}_k$) is the creation (annihilation) operator for acoustic phonons. The energy of the acoustic phonons is described by the Hamiltonian  

\begin{eqnarray}
\hat{H}_{\rm ph} = \sum_{k, \omega_k = v_s |\mathbf{k}|} \hbar \omega_k \hat{b}_k^{\dagger}\hat{b}_k. \label{PhononBath}
\end{eqnarray}

In order to derive the open dynamics, we consider acoustic phonons at thermal equilibrium and weakly coupled to the system. The latter is not a strong assumption and is usually satisfied for color centers in diamond for temperatures ranging from 10 mK to 500 K~\cite{Ariel2018}. After applying the Born and secular approximations, we obtain the following Markovian master equation for the orbital and spin degrees of freedom

\begin{eqnarray} \label{QuantumMasterEquation}
{\mbox{d} \hat{\rho}_{\rm s} \over \mbox{d} t} = {1 \over i\hbar} [\hat{H}_{\rm eff}, \hat{\rho}_{\rm s}] + \sum_{i \neq j}\gamma_{ij}\left[\hat{L}_{ij}\hat{\rho}_{\rm s} \hat{L}_{ij}^{\dagger}-{1 \over 2}\left\{\hat{L}_{ij}^{\dagger}\hat{L}_{ij},\hat{\rho}_{\rm s} \right\} \right],
\end{eqnarray}

where $\hat{L}_{ij} = |i\rangle \langle j|$ defines a set of normalized projection operators, $\ket{i}$ are the normalized eigenstates of the effective Hamiltonian~(\ref{EffectiveHamiltonian}), and $\gamma_{ij}$ are the phonon relaxation rates associated with the transition $\ket{i} \rightarrow \ket{j}$. In the basis $|i \rangle$, the off-diagonal matrix elements can be analytically solved, and are given by $\rho_{ij}(t) = \rho_{ij}(0)\mbox{exp}\left[-\left( \mbox{i} \omega_{ij} + \Gamma_{ij} \right)t \right]$ for $i \neq j$, where $\omega_{ij} = (E_i - E_j)/\hbar$ and $\Gamma_{ij}$ is the effective relaxation rate

\begin{equation}\label{Gamma}
\Gamma_{ij} = {1 \over 2}\left(\sum_{k \neq i} \gamma_{ki} + \sum_{k \neq j}\gamma_{kj} \right).
\end{equation}

To evaluate the individual rates we use the decomposition $\gamma_{ij} = \gamma_{ij}^{\rm 1-ph}+\gamma_{ij}^{\rm 2-ph}$, where $\gamma_{ij}^{\rm 1-ph}$ and $\gamma_{ij}^{\rm 2-ph}$ are the relaxation rates induced by one- and two-phonon processes, respectively. Now, we discuss the low and high temperature regimes for the particular case of color centers. At low temperatures, $T < 10$ K, acoustic phonons dominate the thermal dependence of $\gamma_{ij}$, and therefore $\gamma_{ij} \approx \gamma_{ij}^{\rm 1-ph}$. In such a case, acoustic phonons can be described by a smooth phononic spectral density function, which scales as $\sim \omega^3$ in a three-dimensional lattice. By employing the Fermi golden rule theory to first-order using the linear contribution of the electron-phonon interaction~(\ref{eph}), we obtain the following expression for the one-phonon relaxation rates 

\begin{eqnarray}
\gamma_{ij}^{\rm 1-ph} = \left\{ \begin{array}{lcc}
                       {\displaystyle{\omega_{ji}^3} \over \displaystyle{\pi \hbar v_{\rm s}^3}} \Omega  E n(\omega_{ji}) & \mbox{if}  & E_j > E_i \\  
  {\displaystyle{\omega_{ij}^3} \over \displaystyle{\pi \hbar v_{\rm s}^3}}\Omega  E  \left[n(\omega_{ji})+1\right] & \mbox{if}  & E_i < E_j  
                      \end{array} \right.,
\end{eqnarray}

where $\Omega$ is the volume of the unit cell in a diamond lattice, $E = (F_{\rm acous}/\omega_{\rm ph})^2/(2\mu)$, $F_{\rm acous}$ is the coupling constant associated with acoustic phonons, and $n(\omega) = [\mbox{exp}(\hbar \omega / k_{\rm B}T)-1]^{-1}$ is the mean number of phonons at thermal equilibrium. The expression for $\gamma_{ij}^{\rm 1-ph}$ has the same structure as the photon-induced relaxation rates of the Wigner-Weisskopf theory of the spontaneous emission. In our case, the rate $\Gamma_{ij}$~(\ref{Gamma}) will contribute to both absorption and emission processes, leading to a linear thermal scaling at high temperatures, \textit{i.e.}, $\Gamma_{ij} \propto T$~\cite{Janke2015} within our temperature range of analysis. \par

At high temperatures, $T > 100$ K, experimental measurements of the longitudinal relaxation time $T_1$ with negatively charged nitrogen-vacancy~\cite{Jarmola2012} and neutral silicon-vacancy~\cite{Green2017} centers shows that two-phonon processes are composed by two important contributions: i) Raman processes leading to $1/T_1 \propto T^n$ ($n=5$ for NV$^{-}$ and $n=7$ for the SiV$^{0}$) and ii) Orbach-type processes leading to $1/T_1 \propto [\mbox{exp}(\Delta E/k_B T)-1]^{-1}$, where $\Delta E$ is the energy of a phonon ($\Delta E = 73$ meV for NV$^{-}$ and $\Delta E = 22$ meV for SiV$^{0}$). However, recent experimental measurements for the SiV$^{-}$ show that in the temperature region between 500 mK and 2.3 K there is a more involved temperature dependence for $1/T_1$, which is beyond of the scope of this work. At room temperatures, the Raman processes are the most dominant contribution~\cite{Jarmola2012,Green2017}, which allows us to neglect the Orbach rates in a phenomenological model at high temperatures.\par

We can address the problem of modelling the contribution of two-phonon processes by considering a rate described by $\gamma_{ij}^{\rm 2-ph} = A_{ij}T^{n}$, where $A_{ij}$ is a fit parameter and $n = 5,\; 7$ for the NV$^{-}$, SiV$^{0}$ center, respectively. A detailed derivation of the term $A_{ij}T^{n}$ is presented in \ref{AppendixC}. Also, we note that the observed $T^3$ law for the phonon-induced Raman processes of the SiV$^{-}$~\cite{Janke2015} is only valid for the electronic transition between excited and ground states, as there only phonon modes with energy higher than the spin-orbit coupling contribute to the Raman transition. This case is crucially different from the internal ground dynamics presented in this work. We remark that a full derivation of the Orbach rates is beyond the scope of this paper, and we do not consider it since we are interested in the phenomenology thermal response of the ESR signal. In the next section, we introduce the analytical expression for the ESR absorption spectrum derived from the linear response theory.

\section{ESR absorption spectrum} \label{ESR absorption spectrum}

The response of the system when the oscillating magnetic field is swept across spin-flip resonance transitions depends on the intensity of the interaction Hamiltonian $\hat{V}(t) = \gamma_{\rm s} b_1 \hat{S}_{\rm x} \cos(\omega t)$. Nevertheless, in the weak driven limit, \textit{i.e.}, when $|\gamma_{\rm s} b_1 | \ll \mbox{max}|\hat{H}_{\rm eff}|$ (where $\hat{H}_{\rm eff}$ is the effective Hamiltonian~(\ref{EffectiveHamiltonian})), the absorption spectrum can be defined as the imaginary part of the dynamical susceptibility~\cite{Mahan2000, Tokmakoff2009}

\begin{equation}
I(\omega) = \mbox{Im}\left(\int_{0}^{\infty} e^{i\omega t} \langle \hat{S}_{\rm x}(t) \hat{S}_{\rm x}(0) \rangle_{\rm ss}  \, dt\right), 
\end{equation}

where $\hat{S}_{\rm x}(0)$ is the initial $x$-component of the spin operator and the expectation value $\langle \hat{S}_{\rm x}(t) \hat{S}_{\rm x}(0) \rangle_{\rm ss} = \Tr(\hat{S}_x(t) \hat{S}_x(0) \hat{\rho}_{\rm ss})$ is calculated using the steady state $\hat{\rho}_{\rm ss}$ obtained from the quantum master equation~(\ref{QuantumMasterEquation}). It is convenient to write the spin operator 
$\hat{S}_{\rm x}(0)$ in the basis spanned by the eigenstates of the effective Hamiltonian~(\ref{EffectiveHamiltonian}), obtaining

\begin{eqnarray}
\hat{S}_{\rm x}(0) = \sum_{i,j}g_{ij}|i\rangle\langle j|, 
\quad \quad  \hat{H}_{\rm eff} = \sum_{i} E_i \ket{i}\bra{i},
\end{eqnarray}

where $g_{ij} =  \langle i | \hat{S}_x |j\rangle$. The time-dependent spin operator $\hat{S}_{\rm x}(t)$ is calculated using the following quantum dynamical map 

\begin{eqnarray}
\hat{S}_{\rm x}(t) = \Lambda_t \hat{S}_{\rm x}(0) = \sum_{i,j}g_{ij} \left(\Lambda_t|i\rangle\langle j| \right) = \sum_{i,j} g_{ij} \hat{\rho}_{\rm s}^{(i,j)}(t),
\end{eqnarray}

where $\hat{\rho}_{\rm s}^{(i,j)}(t)$ is the solution of the master equation~(\ref{QuantumMasterEquation}) starting from the initial condition $\hat{\rho}_{\rm s}(0) = |i\rangle \langle j|$. By calculating the expectation value $\langle \hat{S}_{\rm x}(t) \hat{S}_{\rm x}(0) \rangle_{\rm ss}$ using the steady state $\hat{\rho}_{\rm ss} = \mbox{exp}(-\beta \hat{H}_{\rm eff})/\mbox{Tr}(\mbox{exp}(-\beta \hat{H}_{\rm eff}))$, we get the following analytical expression for the ESR absorption spectrum 

\begin{eqnarray} \label{ESR-Spectrum}
I(\omega) = \sum_{i\neq j}|g_{ij}|^2 p_i(T) {(\Gamma_{ij}/2) \over (\omega-\omega_{ij})^2+(\Gamma_{ij}/2)^2}.
\end{eqnarray}

The ESR absorption spectrum is a sum of Lorentzian functions with peaks at the resonant frequencies $\omega_{ij} = (E_i-E_j)/\hbar$ obtained from the effective Hamiltonian~(\ref{EffectiveHamiltonian}) and broadened by the phonon relaxation rates $\Gamma_{ij}$ given in equation~(\ref{Gamma}). The Boltzmann distribution $p_i(T)$ is given by

\begin{eqnarray}
P_i(T) = e^{- \beta E_i}\left(\sum_{i}e^{- \beta E_i} \right)^{-1}, \quad \quad \beta = {1 \over k_{\rm B}T}.
\end{eqnarray}

In the next section, we study the combined effect of the Ham reduction factor $p_{\rm acous}(T)$ and the phonon relaxation rates $\Gamma_{ij}$ on the ESR absorption spectrum. \par

\section{Motional and dynamical suppression effects on the ESR absorption spectrum}

\begin{figure}[htb]
\centerline{\includegraphics[width=.8\textwidth]{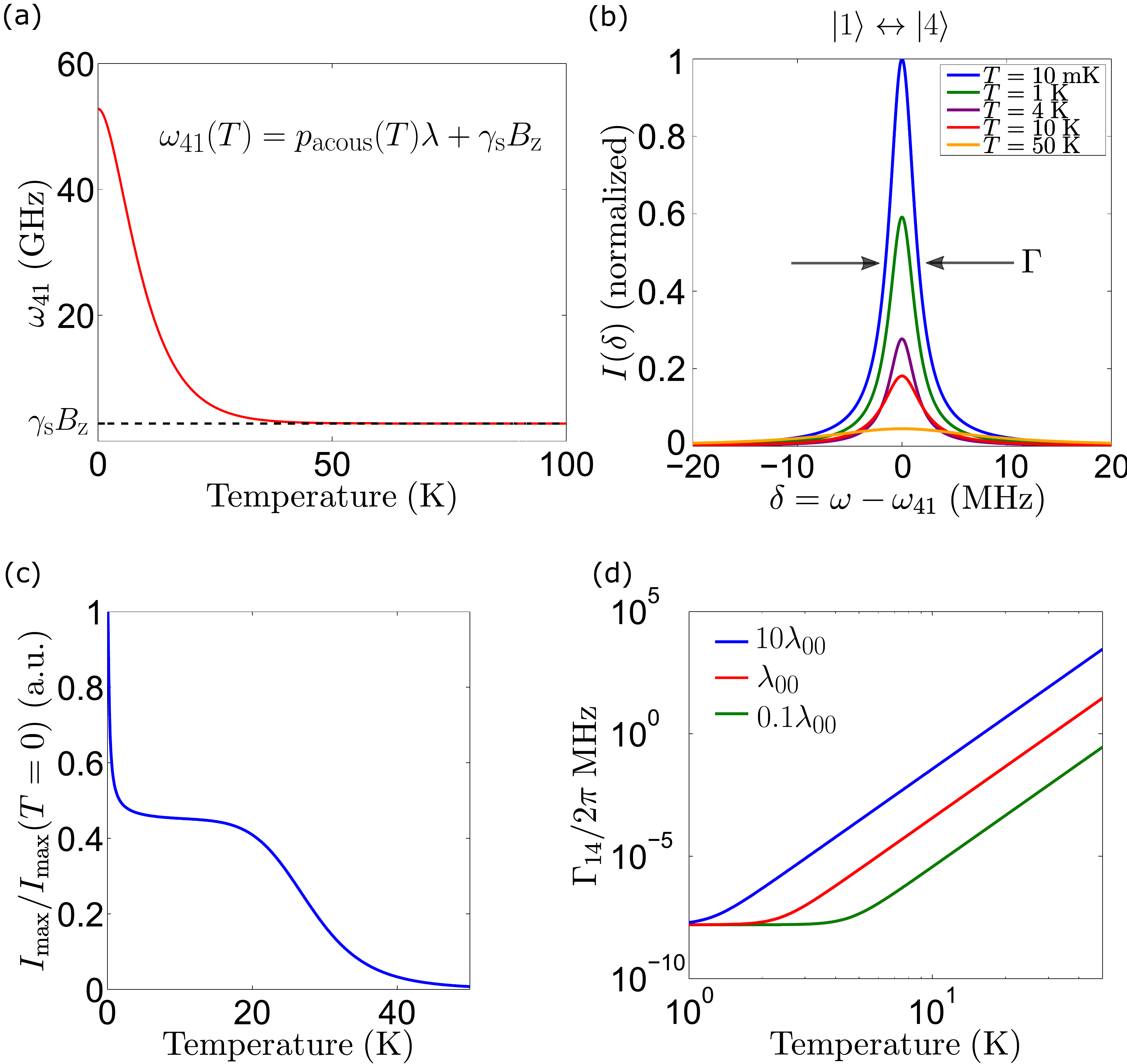}}
\caption{(a) Thermal dependence of the resonant frequency $\omega_{41} = p_{\rm acous}(T)\lambda + \gamma_{\rm s}B_{\rm z}$, where the dashed line is the value $\gamma_{\rm s}B_{\rm z}$. (b) ESR absorption spectrum $I(\delta)$ associated with the transition $|1\rangle \leftrightarrow |4\rangle$ calculated in terms of the detuning frequency $\delta = \omega - \omega_{14}$ and for different temperatures. We use a static magnetic field with components $B_{\rm x} = B_{\rm y} =0$ and $B_{\rm z} = 1000$ G, a spin-orbit coupling $p_{\rm acous}(0)\lambda = 50$ GHz, and a local strain $\gamma_{\rm x} = \gamma_{\rm y} = 1$ GHz. (c) Maximum value of the ERS signal in comparison with its value at zero temperature $I_{\rm max}/I_{\rm max}(T=0)$. (d) Effective relaxation rate $\Gamma_{14}$ in logarithmic scale as a function of temperature and for different two-phonon coupling constants $\lambda_{00} \approx \lambda_{00ij}$, see Appendix~\ref{AppendixC}. The coupling constant $\lambda_{00}$ is taken equal to $2.4$ $\upmu$eV in order to obtain the same temperature dependence of the neutral silicon-vacancy center \cite{Green2017}.} \label{figure4}
\end{figure}

In order to better illustrate the combined thermal effects induced by the Ham reduction factor $p_{\rm acous}(T)$ and the phonon relaxation rates $\Gamma_{ij}$ on the ESR absorption spectrum we analyse the individual transition between two eigenstates of the effective Hamiltonian~(\ref{EffectiveHamiltonian}). Let us consider the transition $|1\rangle \leftrightarrow |4\rangle$, where $\ket{1}$ and $\ket{4}$ are the lowest and highest energy levels of the effective Hamiltonian, respectively. For the particular case of color centers in diamond, the spin-orbit coupling constant is larger than the strain parameters, and therefore we can assume that $\lambda \gg (\gamma_x^2+\gamma_y^2)^{1/2}$. In such a case, the resonant frequency is given by $\omega_{41} \approx p_{\rm acous}(T) \lambda + \gamma_{\rm s} B_{\rm z}$, where $\mathbf{B} = B_{\rm z} \hat{z}$ is the static magnetic field. At high temperatures we have $p_{\rm acous}(T) \approx 0$ (see Figure~\ref{figure3}) leading to $\omega_{41} \approx \gamma_{\rm s} B_{\rm z}$, which is a drastic reduction of the resonant frequency induced by the quenched spin-orbit interaction. In figure~\ref{figure4}-(a) we plot the frequency $\omega_{41} = p_{\rm acous}(T)+\gamma_{\rm s} B_{\rm z}$ as a function of temperature. At zero temperature, we have $\omega_{41}  \approx 50$ GHz, however when the temperature increases the resonant frequency is fully determined by the Zeeman energy term $\gamma_{\rm s} B_z$. Therefore, the peak of the ESR signal around $\omega_{41}$ moves as a function of temperature as shown in figure~\ref{figure4}-(a). This motional effect of the ESR signal is a consequence of the strong JT interaction, which is encapsulated in the Ham reduction factor. \par

In figure~\ref{figure4}-(b) we plot the ESR signal for the transition $|1\rangle \leftrightarrow |4\rangle$ in terms of the detuning $\delta = \omega-\omega_{41}$ and considering $\mathbf{B} = (0,0,1000)$ G, $p_{\rm acous}(0)\lambda = 50$ GHz and $\gamma_{\rm x} = \gamma_{\rm y} = 1$ GHz, and a phenomenological magnetic noise $\Gamma_{\rm mag} = 3$ MHz. By looking the shape of the ESR absorption spectrum we confirm that the intensity is drastically suppressed at high temperatures. This effect can be explained in terms of the phonon relaxation rate $\Gamma_{14}$ and the Boltzmann distribution factor $p(T)$ introduced in equation~(\ref{ESR-Spectrum}). From equation~(\ref{ESR-Spectrum}), we deduce that the intensity and the full width at half maximum (FWHM) of the ESR signal are given by $\Gamma$ and $2|g_{14}|^2 p_1(T)/\Gamma$, respectively at the resonant frequency $\omega = \omega_{41}$. The phonon relaxation rate is given by $\Gamma =  \Gamma_{\rm mag} + \Gamma_{14}$, where the constant magnetic noise $\Gamma_{\rm mag}$ describes the effect of magnetic impurities in the lattice and $\Gamma_{14}$ is calculated from equation~(\ref{Gamma}). Because of the dominant contribution of two-phonon processes at high temperatures, we obtain a temperature scaling $T^7$ for the relaxation rate $\Gamma_{14}$, which is shown in figure~\ref{figure4}-(d) for different two-phonon coupling constants $\lambda_{00} \approx \lambda_{00ij}$ (see~\ref{AppendixC} for further details). As a consequence, the strong thermal dependence $\Gamma_{41} \propto T^7$ can explain the reduction of the ESR signal when temperature increases and its absence at high enough temperatures. \par

In addition, we plot the intensity of the ESR signal at the resonant frequency $\omega = \omega_{41}$ in figure~\ref{figure4}-(c). We numerically confirm that the intensity is reduced to zero above temperatures $T > 100$ K. In other words, the suppression of the ESR signal at room temperature is because of the activation of two-phonon processes and not for the effect of the Ham reduction factors induced by the strong JT effect. Towards this direction, recent experimental observations of the ESR transitions associated with the SiV$^{-}$ centers showed that the signal is observed at 100 mK~\cite{Lukin2017} and also at 4 K~\cite{Pingault2017} but not at room temperature~\cite{Edmonds2008}. Our model predicts that at low temperatures, two-phonon processes are not activated, resulting in an ESR signal with a high intensity, as shown in figure~\ref{figure4}-(b). We remark that other similar systems may not present these motional and suppression effects at high temperatures if the set of parameters $\lambda, \gamma_{\rm x,y}, \Gamma_{ij}$ and $p_{\rm acous}(T)$ are different and/or the assumptions presented here are not satisfied for them. For instance, the vanadium in hexagonal silicon-carbide, a defect system with trigonal symmetry, spin-$1/2$, and ground state with dynamic JT effect still shows evidence for an ESR signal above $50$ K~\cite{Schneider1990, Kunzer1993, Kaufmann1996}.

\section{Conclusion}\label{Conclusion}

In summary, we have presented a microscopic model to study the effect of phonons and temperature on the ESR absorption spectrum in a generic $E \otimes e$ Jahn-Teller system with residual spin $S=1/2$. We derived the effective dynamics for the orbital and spin degrees of freedom by finding an analytical expression for the thermal dependence of the Ham reduction factor $p$. Furthermore, we included one- and two-phonon processes into the relaxation dynamics to analyse thermal effects on the ESR signal. Using our model and the SiV$^{-}$ parameters, we presented a physical explanation for the absence of contrast in the ESR response at high temperatures, and also we show that the signal is recovered at temperatures of the order of tens-hundreds mK. Interestingly, the latter is experimentally confirmed by recent experiments with color centers in diamond~\cite{Lukin2017,Becker2017}. Most importantly, we demonstrated that a strong dynamical JT effect modifies the orbital operators in a short time scale, leading to an effective Hamiltonian whose energy levels are dominated by the Zeeman interaction at high temperatures. On the contrary, phonon-induced relaxation processes sharply decrease the amplitude of the ESR signal due to phonon broadening effects. The model can be used to characterize new spin-$1/2$ systems at low temperatures for metrology and quantum information applications, where a strong JT effect is present.

\section{Acknowledgements}

The authors thank Marcus Doherty and Adam Gali for enlightening discussions and helpful insights. AN acknowledges financial support from Universidad Mayor through the Postdoctoral fellowship. J.~R.~M.~ acknowledges support from ANID-Fondecyt 1180673 and ANID-PIA ACT192023, AFOSR grant FA9550-18-1-0513 and ONR grant N62909-18-1-2180. C.~B. acknowledges funding by Deutsche Forschungsgemeinschaft (DFG) (FOR1493,BE2306/7-1).

\appendix

\section{Ham reduction factors} \label{AppendixA}

In order to introduce the transformation given in equation~(\ref{Transformation}) we first describe the effect of two independent modes $\hat{Q}_{\rm x}$ and $\hat{Q}_{\rm y}$ and after that we extend our results to a continuum of $e$-phonon modes. Firstly, let us consider the following unitary transformation

\begin{equation}
e^{i\epsilon \hat{T}}H_{\rm sJT}e^{-i\epsilon \hat{T}}=H_{\rm sJT}+ e^{i\epsilon \hat{T}}\left[H_{\rm sJT},e^{-i\epsilon \hat{T}}\right],
\end{equation}

where $H_{\rm sJT}$ is the JT Hamiltonian for a two vibrational modes, $\hat{T}=\hat{P}_{\rm x}\hat{\sigma}_{\rm z}+\hat{P}_{\rm y}\hat{\sigma}_{\rm x}$ and $\epsilon = -F/\mu\hbar \omega$. This allows us to decouple orbital and momentum operators, and the transformed JT Hamiltonian at first order in $\epsilon$ (weak coupling regime) will be~\cite{Ham1968}

\begin{eqnarray}
\hat{H}_{\rm sJT}' = E_0 + \sum_{i= \rm x,\rm y}\left({\hat{P}_{i}^2 \over 2\mu} + {1 \over 2}\omega^2\hat{Q}_{i}^2\right) - {F^2 \over \mu \hbar \omega^2}\left(3\hbar - 4\hat{L}_z \hat{L}_{\rm z}^{\rm ph}\right).
\end{eqnarray}

However, we must also apply this transformation to the other terms of the Hamiltonian~(\ref{Hamiltonian-of-the-system}), in particular, we have to apply it to terms that depend on orbital operators $\hat{\sigma}_i$ ($i=\rm x,y,z$). After averaging over the phonon bath we obtain

\begin{equation}
\left\langle\hat{\sigma}_{y}\right\rangle=p\left\langle\hat{\sigma}_{y}\right\rangle_{\rm e} \qquad \left\langle\hat{\sigma}_{x}\right\rangle=q\left\langle\hat{\sigma}_{x}\right\rangle_{\rm e} \qquad \left\langle\hat{\sigma}_{z}\right\rangle=q\left\langle\hat{\sigma}_{z}\right\rangle_{\rm e},
\end{equation}

where the subindex $\rm e$ represents an averaging over the electronic part of the states and $p$ and $q$ are the Ham reduction factors~\cite{Ham1968} which satisfy the relation $q=\frac{1}{2}(p+1)$. We get

\begin{equation}
p = 1-2\epsilon^2 \left\langle \left( {\sin \epsilon P \over \epsilon P} \right)^2 (P_{\rm x}^2+P_{\rm y}^2) \right\rangle_{\rm ph},
\end{equation}
where $P^2 = P_{\rm x}^2+P_{\rm y}^2$. Using the trigonometric identity 
$\sin^2x = (1-\cos 2x)/2$, we obtain
 
\begin{equation}
p = 1-\left\langle \left(1-\cos 2\epsilon P \right)\right\rangle_{\rm ph} =\sum_{n=0}^{\infty}{(2\epsilon)^{2n} \over (2n)!}\langle P^{2n} \rangle_{\rm ph}.
\end{equation}

Introducing the quantization of phonons $P_{\rm x,y} = i\sqrt{\mu\hbar \omega/2}(b_{\rm x,y}^{\dagger}-b_{\rm x,y})$ and using the Fock representation $\ket{k} = \ket{k_{\rm x},k_{\rm y}}$, we deduce that

\begin{equation}
P^2 \ket{k} = (P_{\rm x}^2+P_{\rm y}^2)\ket{k_{\rm x},k_{\rm y}} = \mu \hbar \omega (2n(\omega)+1)\ket{k_{\rm x},k_{\rm y}}, \label{P2}
\end{equation}

where $n(\omega) = [\mbox{exp}(\hbar \omega /k_{\rm B} T)-1]^{-1}$ is the mean number of phonons at thermal equilibrium. From Eq.~(\ref{P2}) we get the following relation

\begin{equation}
P^{2n}\ket{k} = \left[\mu \hbar \omega (2n(\omega)+1) \right]^n \ket{k}.
\end{equation}

Now, if we calculate the expectation value using a thermal phonon state described by $\rho_{\rm ph} = e^{-\beta}/Z_{\rm ph}$, where $Z_{\rm ph} = \mbox{Tr}_{\rm ph}(e^{-\beta H_{\rm ph}})$ and $\beta = 1/k_{\rm B} T$, we obtain

\begin{equation}
\langle P^{2n} \rangle_{\rm ph} = {1 \over Z_{\rm ph}}\sum_{k} \bra{k} P^{2n} e^{-\beta H_{\rm ph}} \ket{k} = \left[\mu \hbar \omega (2n(\omega)+1) \right]^n,
\end{equation}

where we have used $H_{\rm ph} = \hbar\omega(b_{\rm x}^{\dagger}b_{\rm x}+b_{\rm y}^{\dagger}b_{\rm y})$ and $e^{-\beta H_{\rm ph}} \ket{k} = e^{-2\beta n(\omega)}\ket{k}$. Finally, the following Ham reduction factor is obtained

\begin{equation}
p = \sum_{n=0}^{\infty}{(2\epsilon)^{2n} \over (2n)!} \left[\sqrt{\mu \hbar \omega (2n(\omega)+1)} \right]^{2n} = 
\cos\left(2\epsilon \sqrt{\mu \hbar \omega \coth\left({\hbar \omega \over 2  k_{\rm B} T }\right)}\right),
\end{equation}

where $2n(\omega)+1 = \coth\left(\hbar \omega / 2 k_B T \right)$. To second order in $\epsilon$ the above expression is reduced to

\begin{equation}
p^{(2)} \approx 1 - 2 \left(F \sqrt{{\hbar \over \mu \omega}}\right)^2 {1 \over (\hbar \omega)^2}\coth\left( {\hbar \omega \over 2 k_B T} \right).
\end{equation}

The generalization of $p^{(2)}$ to a continuum of phonon modes can be made by adding the contribution of all phonon modes, following this we recover our result given in equation~(\ref{p-factor}).

\section{Effective Hamiltonian} \label{AppendixB}
As previously we discussed in~\ref{AppendixA} we first consider the effect of two independent phonon modes. Let us consider the Liouville equation

\begin{equation}
\dot{\hat{\rho}} = {1 \over i \hbar} [\hat{H},\hat{\rho}],
\end{equation}

where $\hat{H} = \hat{H}_{\rm JT} + \hat{H}_{\rm so} + \hat{H}_{\rm z}$ is the total Hamiltonian of the system, where
 
\begin{equation}
\hat{H}_{\rm JT} = E_0 + \left[{1 \over 2\mu}\left(\hat{P}_{\rm x}^2 +\hat{P}_{\rm y}^2 \right) + {1 \over 2}\omega^2\left(\hat{Q}_{\rm x}^2+\hat{Q}_{\rm y}^2 \right) \right] + F (Q_{\rm x} \sigma_{\rm z} - Q_{\rm y} \sigma_{\rm x}),
\end{equation}

is the JT Hamiltonian. We use the same unitary transformation introduced in equation~(\ref{Transformation})
\begin{equation}
\hat{\rho}' = \hat{U} \hat{\rho} \hat{U}^{\dagger}, \quad \quad \hat{U} = \mbox{exp}\left(i \epsilon \hat{T}\right), \quad \quad \hat{T} = (\hat{P}_{\rm x} \sigma_{\rm z} + \hat{P}_{\rm y} \hat{\sigma}_{\rm x}),
\end{equation}
where $\epsilon = -F/\mu\hbar \omega$. Now, we consider the Born or Hartree approximation $\hat{\rho}(t) = \hat{\rho}_{\rm s}(t) \otimes \hat{\rho}_{\rm ph}$, where phonons are considered in thermal equilibrium. Expanding in powers of $\epsilon$, we obtain 

\begin{equation}
\hat{\rho}' \approx \left(1+i\epsilon \hat{T} \right) \hat{\rho} \left(1-i\epsilon \hat{T}\right) \approx \rho + i\epsilon [T,\rho].
\end{equation}

By calculating the reduced density operator $\mbox{Tr}_{\rm ph}(\hat{\rho}')$, we obtain 

\begin{equation}
\mbox{Tr}_{\rm ph}(\hat{\rho}') \approx  \mbox{Tr}_{\rm ph}(\hat{\rho}) + i\epsilon\mbox{Tr}_{\rm ph}([\hat{T},\hat{\rho}]) = \hat{\rho}_{\rm s} + i\epsilon \hat{\rho}_{\rm s} \mbox{Tr}(\hat{T} \hat{\rho}_{\rm ph}) = \hat{\rho}_{\rm s},
\end{equation}

where we have used $\mbox{Tr}(T \rho_{\rm ph}) = 0$ since $\hat{T}$ is a linear operator in $P_{x}$ and $P_y$ over the phonon sub-space. By taking the trace over the phonon degrees of freedom, we deduce
\begin{equation}
\dot{\hat{\rho}}_{\rm s} = {1 \over i\hbar}\mbox{Tr}_{\rm ph}([\hat{H'},\hat{\rho}']),
\end{equation}
where $\mbox{Tr}_{ \rm ph}([\hat{H}',\hat{\rho}']) = \mbox{Tr}_{ \rm ph}(\hat{H}' \hat{\rho}') - \mbox{Tr}_{\rm ph}(\hat{\rho}' \hat{H}') = \mbox{Tr}_{\rm ph}(\hat{H}'\hat{\rho}') - \mbox{H.C.}$. By simplicity we calculate the first term contribution associated with each Hamiltonian

\begin{eqnarray}
\mbox{Tr}_{\rm ph}(\hat{H}_{\rm JT}' \hat{\rho}') &=& \hat{\rho}_{\rm s}\left(1+\langle \hat{H}_{\rm ph} \rangle - { 3F^2 \over \mu \omega^2 } \right), \\
\mbox{Tr}_{\rm ph}(\hat{H}_{\rm so}' \hat{\rho}') &=& p \hat{H}_{\rm so} \hat{\rho}_{\rm s}, \\
\mbox{Tr}_{\rm ph}(\hat{H}_{\rm z}' \hat{\rho}') &=& p \hat{H}_{\rm zo} \hat{\rho}_{\rm s} + \hat{H}_{\rm zs} \hat{\rho}_{\rm s}
\end{eqnarray}

where $p$ is the Ham reduction factor, $\hat{H}_{\rm zo}$ is the orbital Zeeman interaction and $\hat{H}_{\rm zs}$ is the spin Zeeman interaction. Finally, we obtain
\begin{equation}
\dot{\hat{\rho}}_{\rm s} = {1 \over i\hbar}([\hat{H}_{\rm eff},\hat{\rho}_{\rm s}]),
\end{equation}
where $\hat{H}_{\rm  eff} = p \hat{H}_{\rm so} + p \hat{H}_{\rm zo} + \hat{H}_{\rm zs}$ is the effective Hamiltonian.

\section{Two-phonon processes} \label{AppendixC}
To model two-phonon processes, we use the electron-phonon Hamiltonian given in equation~(\ref{eph}). In a three-dimensional lattice without cubic symmetry, we can model the coupling constants in the continuum limit as \cite{Weiss2008}:

\begin{eqnarray}
\lambda_{ij,k} &\rightarrow & \lambda_{ij}(\omega)  =  \lambda_{0ij} \left({\omega \over \omega_{\rm D}} \right)^{1/2}, \\
\lambda_{ij,kk'} &\rightarrow & \lambda_{ij}(\omega,\omega')  =  \lambda_{00ij} \left({\omega \over \omega_{\rm D}} \right)^{1/2} \left({\omega' \over \omega_{\rm D}} \right)^{1/2}, 
\end{eqnarray}

where $\omega_D$ is the Debye frequency in diamond and $\lambda_{0ij}, \lambda_{00ij}$ are coupling constants. To introduce the transition rate between eigenstates of the effective Hamiltonian we will focus on phonon processes satisfying the energy condition $\omega_k-\omega_{k'} = (E_i-E_j)/\hbar$, where $E_i$ are the eigenenergies of the effective Hamiltonian (\ref{EffectiveHamiltonian}). The transition rate between two eigenstates can be defined as~\cite{Ariel2018}

\begin{eqnarray}
\gamma_{i \rightarrow j} &=& \sum_{k: \; \omega_k = v_{\rm s}|\mathbf{k}|} \; \sum_{k': \; \omega_{k'} = v_{\rm s} |\mathbf{k}'|} \Gamma_{i,n_k,n_{k'}}^{j, n_k-1,n_{k'}+1}.
\end{eqnarray}

To second order in perturbation theory the Fermi golden rule states that

\begin{eqnarray}
\Gamma_{i,n_k,n_{k'}}^{j, n_{l},n_{l'}} &=& {2 \pi \over \hbar^2}\left|V_{i,n_k,n_{k'}}^{j, n_{l},n_{l'}} +\sum_{m}\sum_{p,p'}   {V_{j,n_l,n_{l'}}^{m,n_p,n_{p'}}  V_{m,n_p,n_{p'}}^{i,n_k,n_{k'}}\over E_{i,n_k,n_{k'}} - E_{m,n_p,n_{p'}} }\right|^2 \nonumber \\ 
&& \hspace{0.5 cm} \times \delta(\omega_{ji}+ n_l \omega_l + n_{l'}\omega_{l'}- n_k \omega_k - n_{k'}\omega_{k'}),  \label{FermiGoldenRule}
\end{eqnarray}

where $V_{i,n_k,n_{k'}}^{j, n_{l},n_{l'}} = \langle i,n_k,n_{k'} | \hat{H}_{\rm e-ph} | j, n_{l},n_{l'}\rangle$, $|n_k\rangle$ represents a Fock state for the $k$-th vibrational mode and $E_{i,n_k,n_{k'}}= \hbar(\omega_i+n_k \omega_k + n_{k'}\omega_{k'})$, with $E_i = \hbar \omega_i$. In the continuum limit the density of states (DOS) due to acoustic phonons is given by

\begin{eqnarray}
\mathcal{D}(\omega) &=& \Omega\int_{}^{}\frac{d^3 k}{(2\pi)^3} \delta(\omega - v_{\rm s}|\mathbf{k}|)\nonumber\\
&=& \frac{\Omega}{(2\pi)^3}\int_{} d\Omega \int_{0}^{k_{\rm D}} dk\, k^{2} \delta(\omega - v_{\rm s} k)\nonumber\\
&=& D_0 \omega^2 \Theta(\omega_{\rm D} - \omega). 
\label{DOS-d}
\end{eqnarray}

where $d\Omega = \sin \theta d\theta d\phi$ is the differential solid angle, $\omega_{\rm D} = v_{\rm s} k_{\rm D}$, $D_0 = \Omega /(2\pi^2 v_{\rm s}^3)$ and $\Theta(x)$ is the Heaviside function. Using the DOS and the intermediate phonon states $\ket{n_p,n_{p'}} = \{\ket{n_k-1,n_{k'}},\ket{n_k,n_{k'}+1}\}$, we obtain the following transition rate

\begin{eqnarray}
\gamma_{i \rightarrow j} &=& a_{ij}T^5+b_{ij}T^6 + c_{ij}T^7.
\end{eqnarray}

Each coefficient is defined as

\begin{eqnarray}
a_{ij} &=& {2 \pi D_0^2 k_{\rm B}^5 \over \hbar^5(\hbar \omega_{\rm D})^2}
\int_{0}^{x_{\rm D}} f_{ij}(x)\left|\sum_{m}\lambda_{0jm}\lambda_{0mi}
                       g_{ijm}(x) \right|^2 dx, \\
b_{ij} &=&  {8 \pi D_0^2 k_{\rm B}^6 \over \hbar^6 (\hbar \omega_{\rm D})^2}
\int_{0}^{x_{\rm D}} f_{ij}(x) \mbox{Re}\left\{\lambda_{00ij}^{\ast}\sum_{m}\lambda_{0jm}\lambda_{0mi}
                        g_{ijm}(x) \right\} \, dx,  \\                     
c_{ij} &=& {8 \pi D_0^2 |\lambda_{00ij}|^2 k_{\rm B}^7 \over \hbar^7 (\hbar \omega_{\rm D})^2}
\int_{0}^{x_{\rm D}} f_{ij}(x) \, dx,
\end{eqnarray}

where $x_{\rm D} = \hbar \omega_{\rm D}/k_{\rm B}T$ and

\begin{eqnarray}
f_{ij}(x) &=& x^3(x-x_{ji})^3n(x)[n(x-x_{ji})+1], \\
g_{ijm}(x) &=& {1 \over x_{im}+x}+{1 \over x_{jm}-x},
\end{eqnarray}

where $n(x) = [\mbox{exp}(x)-1]^{-1}$ and $x_{ij} = \hbar \omega_{ij}/(k_{\rm B} T)$. By considering that the coupling with two phonons is described by a real parameter $\lambda_{ij kk'} \in \mathds{R}$, we conclude that $b_{ij} = 0$. In such case, we have $\gamma_{i \rightarrow j} = a_{ij}T^5 + c_{ij}T^7$. For a single SiV$^{-}$ center under the presence of a static magnetic field $\mathbf{B} = B_{\rm z} \hat{z}$, the allowed transitions induced by phonons are given by $|1\rangle \leftrightarrow |3\rangle$ and $|2\rangle \leftrightarrow |4\rangle$. Therefore, the only non-zero linear electron-phonon parameters are $\lambda_{042}$, $\lambda_{024}$, $\lambda_{031}$ and $\lambda_{013}$. As a consequence, we obtain $a_{i\rightarrow j}^{\rm SiV}  = 0$ due to the fact that $\sum_{m}\lambda_{0jm}\lambda_{0mi} = 0$ for every $i,j = 1,2,3,4$. Now, we assume that $\lambda_{00 ij} \approx \lambda_{00}$. Second, due to the Ham reduction factors the terms $x_{ij} = \hbar \omega_{ij}/(k_{\rm B}T) \approx 0$ for $T > 100$ K and low magnetic fields. Therefore, we obtain the following contribution to the effective transition rate $\Gamma_{ij}$ (see equation~(\ref{Gamma}))

\begin{equation}
\Gamma_{ij} \approx {10380 \pi D_0^2 |\lambda_{00}|^2 k_{\rm B}^7 \over \hbar^7 (\hbar \omega_{\rm D})^2} \times T^7,
\end{equation}

where we have numerically solved the integral $I =\int_{0}^{x_{\rm D}} f(x) \, dx $ for $f(x) = x^6 n(x)[n(x)+1]$, obtaining a constant value $I = 432.487$ in the temperature interval $(0-300)$ K. In order to compare our model to real systems, we extract the experimental value $1/T_1 = A_{\rm Raman}T^7$ for the neutral silicon-vacancy center, where $A_{\rm Ramam} = 5.0 \times 10^{-13}$ s$^{-1}$K$^{-7}$ \cite{Green2017}. By assuming that $\lambda_{00ij} \approx \lambda_{00}$, we estimate $\lambda_{00}$ to be approximately $2.4$ $\upmu$eV.

\section*{References}

\bibliographystyle{unsrt}

\end{document}